\makeatletter
\@ifundefined{@parse@version@dash}{%
\def\@parse@version#1{\@parse@version@0#1}
\def\@parse@version@#1/#2/#3#4#5\@nil{%
\@parse@version@dash#1-#2-#3#4\@nil}
\def\@parse@version@dash#1-#2-#3#4#5\@nil{%
  \if\relax#2\relax\else#1\fi#2#3#4 }
}{}

\documentclass[twocolumn,aps,pre,groupaddress]{revtex4-2}

\usepackage{graphicx}
\usepackage{amsmath,amssymb}
\usepackage[]{natbib}
\usepackage{xcolor}
\usepackage{dcolumn}
\usepackage{bm}
\usepackage[mathlines]{lineno}

\begin{document}

\title{Violent relaxation in one-dimensional self-gravitating system: deviation from the Vlasov limit due to finite-$N$ effect}

\author{Tirawut Worrakitpoonpon}
\email[Corresponding author: ]{worraki@gmail.com}
\affiliation{Institute of Science, Suranaree University of Technology, Nakhon Ratchasima 30000, Thailand}
\date{}

\begin{abstract}
We investigate the effect of a finite particle number $N$ on 
the violent relaxation leading to the Quasi-Stationary State (QSS)
in a one-dimensional self-gravitating system.
From the theoretical point of view, we demonstrate that
the local Poissonian fluctuations embedded in the initial state
give rise to an additional term proportional to $1/N$
in the Vlasov equation. This term designates the strength of
the local mean-field variations by fluctuations.
Because it is of the mean-field origin, we interpret it differently
from the known collision term in the way that it
effects the violent relaxation stage.
Its role is to deviate the distribution function  
from the Vlasov limit, in the collisionless manner,
at a rate proportional to $1/N$ while the 
violent relaxation is progressing. This hypothesis is tested
by inspecting the QSSs in simulations of various $N$.
We observe that the core phase-space density 
can exceed the limiting density deduced from the Vlasov
equation and its deviation degree is in accordance
with the $1/N$ estimate.
This indicates the deviation from the 
standard mean-field approximation of the violent relaxation process
by that $1/N$ term.
In conclusion, the finite-$N$ effect has a significant contribution
to the QSS apart from that it plays
a role in the collisional stage that takes place long after.
The conventional collisionless Vlasov equation might not be able 
to describe the violent relaxation of a system of particles properly
without the correction term of the local finite-$N$ fluctuations.
\end{abstract}


\maketitle

\section{Introduction} \label{intro}
Long-range interacting systems 
constitute a major branch in statistical mechanics and 
attract much interest due to their complex nature
(see \cite{campa_etal_LRreview_2009} for review).
In general, an out-of-equilibrium state
evolves first through a stage of violent relaxation in which
the dynamics is governed by a self-consistent mean field
with a time scale of the order of dynamical times. 
This stage is otherwise known as the collisionless evolution 
due to the lack of the collisional interaction between the particles.
That collisionless dynamics during that stage 
can be described by the Vlasov equation.
At the end of the violent relaxation, 
the system settles into a the Quasi-Stationary State (QSS).
A descriptions of the violent relaxation and
the resulting QSS have been proposed by 
Lynden-Bell (hereafter LB) \cite{lynden_bell_1967},
initially aiming to explain the inconsistency between the
light distribution of an elliptical galaxy and
the isothermal sphere profile that is the 
thermal equilibrium analogue for the three-dimensional
self-gravitating system. 
That work estimated the violent relaxation time scale 
by the rate of change of the mean field 
and proved that the time scale was of the order of
dynamical times regardless of the system's physical parameters.
This allows a galaxy to be established as a QSS within
a time scale shorter than the age of the Universe.
The statistical description of the 
QSS differs from the thermal 
equilibrium which represents a state of
maximized Boltzmann entropy. Since then, 
the LB theory has became a central pillar 
in the field of astrophysics and the studies of systems governed
by long-range interactions.

While the Vlasov equation addresses the collisionless evolution 
of the long-range interacting systems in the fluid limit, 
the violent relaxation is mostly investigated using systems of particles
in which the time integration of the particle dynamics is simple
compared to the direct integration of the Vlasov equation.
However, the Poissonian fluctuations embedded in particle systems
must be properly considered. A finite particle number $N$ 
constitutes the collision term proportional to $1/N$ 
that arises from the two-point correlation function
in the BBGKY hierarchy
\cite{braun+hepp_1977,chavanis_2010_jstat,fouvry+bar_or_2018}.
It introduces a collisional effect 
to the QSS and drives it to thermal equilibrium with
a time scale increasing with $N$ as verified in
the one-dimensional self-gravitating system
\cite{miller_1996thermal,tsuchiya_et_al_1996,joyce+worrakitpoonpon_2010}
and the Hamiltonian Mean-Field (HMF) model 
\cite{yamaguchi_2003,jain_et_al_2007}. 
Although the knowledge of the finite-$N$ effect on
the long-term evolution to the thermal equilibrium
is well established theoretically and numerically,
its impact on the preceding violent relaxation
is not well investigated. Studies of the HMF model 
have mentioned the finite-$N$ effect on the QSS with 
a deviation of the QSS from the prediction based on
the fluid limit 
\cite{yamaguchi+barre+bouchet+dauxois+ruffo_2004,antoniazzi+califano+fanelli+ruffo_2007,campa_chavanis_2017,santini_et_al_2022}.
The finite-$N$ effect has also been addressed
for a system with attractive power-law interaction
and the correction term to the Vlasov equation
was required in order to describe the particle dynamics
properly \cite{gabrielli+joyce+morand_2014}.
The $N$-dependence of the QSS is interesting 
since it was not described by the original
LB theory which considered the fluid analogue.
Although the finite-$N$ effect was not taken into account,
the LB theory was shown to be applicable in
a limited number of situations. In most cases, the systems underwent
incomplete relaxation that prevented them to achieve the LB configuration
\cite{yawn+miller_1997,yawn+miller_2003,arad+lynden_bell_2005,yamaguchi_2008,joyce+worrakitpoonpon_2011}.

The main purpose of this study is 
to better understand, theoretically and numerically, 
the involvement of the finite-$N$ effect in the violent
relaxation and the resulting QSS in the one-dimensional 
self-gravitating system. The article is organized as follows. 
In Sec. \ref{vlasov_analysis}, the analysis of the Vlasov
equation that incorporates the local Poissonian fluctuations 
is performed and the relevance to the finite-$N$ QSS 
is discussed.
Sec. \ref{ic_unit_sim} details of the 
initial condition for the simulations, the units, and 
the simulation method. In Sec. \ref{nume},
the numerical results and the comparison with
the theoretical framework are presented.
Finally, Sec. \ref{conclusion} concludes this study.

\section{Analysis of the Vlasov equation with local fluctuations} 
\label{vlasov_analysis}
We recall the Vlasov equation that governs 
the collisionless evolution of the distribution function 
$f(\mathbf{x},\mathbf{v})$
\begin{equation}
\frac{df}{dt} = \frac{\partial f}{\partial t}+
\mathbf{v}\cdot\frac{\partial f}{\partial\mathbf{x}}+
\mathbf{a}\cdot\frac{\partial f}{\partial\mathbf{v}} = 0
\label{non_perturbed_dfdt}
\end{equation}
where $\mathbf{a}$ is the self-consistent mean-field acceleration,
an acceleration from all phase-space elements of the system,
derived from the gradient of the mean-field potential $\Phi$ as
\begin{equation}
\mathbf{a}(\mathbf{x}) = -\nabla_{\mathbf{x}}\Phi.
\label{a_gradientphi}
\end{equation}
By definition, $\Phi$ can be written in terms of
the integral of $f$ as
\begin{equation}
\Phi(\mathbf{x}) = \int\int\varphi(\mathbf{x},\mathbf{x'})f(\mathbf{x'},\mathbf{v'})
d\mathbf{x'}d\mathbf{v'}
\label{a_meanfield}
\end{equation}
where $\varphi (\mathbf{x},\mathbf{x'})$ is the pair potential
at $\mathbf{x}$ from the mass element at $\mathbf{x'}$.
According to the LB analysis
\cite{lynden_bell_1967}, there are two types of system
distribution functions: the fine-grained and the coarse-grained
distribution functions. The former one represents
the phase-space density of a system segmented into
smallest divisible phase-space elements, or microcells,
each of which is so small that we can approximate its density
to be uniform and non-zero inside when occupied and
zero otherwise. The latter one is the average of
the phase-space density inside a macrocell, a cell that consists of
multiple microcells. The statistical equilibrium
is attainable on the coarse-grained level, whereas the fine-grained
level never attains because fine-grained entropy does not increase.
Throughout this section, all distribution
functions are the coarse-grained ones.
In simulation aspect, the measured distribution function
and the other profiles derived from it can be considered as 
coarse-grained properties as they
are computed with a finite bin size.

If $f(\mathbf{x},\mathbf{v})$ denotes the point-wise
one-particle distribution function, Eq. (\ref{non_perturbed_dfdt})
is known as the Klimontovich equation. 
In the case of Newtonian gravity,
Eq. (\ref{non_perturbed_dfdt}) is coupled with the Poisson equation
and these equations are called
collectively the Vlasov-Poisson equations.
For an initial state consisting of $N$ particles,
we express the system distribution function $f$ in the form
\begin{equation}
f=\bar{f}+\delta f, 
\label{f_decompos}
\end{equation}
where the first term corresponds to the distribution function
in fluid limit and the second term represents
the local variation due to the Poissonian fluctuations
(see \cite{campa_etal_LRreview_2009} for review).
The variation term vanishes by ensemble average,
i.e., $<\delta f>=0$, so $<f>$ converges to $\bar{f}$.
In statistical sense, the ensemble average of
a quantity or a function can be obtained by averaging it
over infinite number of realizations
starting with initial conditions sampled from the same
distribution function.
In practice, the ensemble average over hundreds of
realizations is found sufficient for reliable
distribution functions of the QSSs.
The fluctuations of $f$ lead to the fluctuations of the 
mean-field acceleration $\delta\mathbf{a}$ accordingly,
i.e., $\mathbf{a}=\mathbf{\bar{a}}+\delta \mathbf{a}$
where $\mathbf{\bar{a}}$ is the mean-field acceleration 
from $\bar{f}$.
From Eqs. (\ref{a_meanfield}) and (\ref{f_decompos}), 
we can deduce that $\delta \mathbf{a}$ is 
an integral of $\delta f$
which vanishes after averaging in the same way as $\delta f$.
All variation terms are scaled by $1/\sqrt{N}$ as they
originate from the Poissonian noise. By substituting
the perturbed quantities into Eq. (\ref{non_perturbed_dfdt})
and performing the ensemble average, the perturbed
Vlasov equation becomes
\begin{equation}
\bigg<\frac{d\bar{f}}{dt}\bigg> = 
-\bigg<\delta\mathbf{a}\cdot\frac{\partial \delta f}{\partial\mathbf{v}}\bigg>
\sim \frac{1}{N},
\label{dfdt_perturbed}
\end{equation}
where
\begin{equation}
\frac{d\bar{f}}{dt}=
\frac{\partial \bar{f}}{\partial t}+
\mathbf{v}\cdot\frac{\partial \bar{f}}{\partial\mathbf{x}}+
\mathbf{\bar{a}}\cdot\frac{\partial \bar{f}}{\partial\mathbf{v}}.
\label{def_dfsdt}
\end{equation}
Although the ensemble average eliminates all terms involving  
first-order fluctuations in Eq. (\ref{dfdt_perturbed}),
a non-vanishing second-order term proportional to $1/N$ remains,
which is of the mean-field origin and embedded in the initial state.
This analytical result may appear at first analogous to 
the one obtained
from the quasi-linear analysis from which the $1/N$ term
is recognized as the collision term
\cite{benetti+marcos_2017,chavanis_2022PhyA}.
In this work, however, we interpret the role of this term and its 
intrinsic properties differently.
Conventional analysis expresses the two
perturbative terms on the right-hand side in forms of
the spatio-temporal Fourier transforms of $\delta f$, 
and the ensemble average of their product corresponds to
the spatio-temporal two-point correlation function. 
This is the result derivable from the standard
Lenard-Balescu equation \cite{chavanis_serie_3}.
We have an alternative view on that term starting from a
simple dynamical perspective that $\delta\mathbf{a}$
which involves the integral of $\delta f$ over the entire
phase space, designates how large the local fluctuation of
mean-field acceleration is at any location.
The product $\delta f\delta\mathbf{a}$ in that term
can then be regarded as the local fluctuations of accelerations
amplified by local variations $\delta f$, which
can also be interpreted as the strength
of the deviation from mean-field approximation.
As the seed of this term, namely $\delta\mathbf{a}$,
is of mean-field origin, we hypothesize that it is associated with
the collisionless relaxation to the QSS rather than
the collisional relaxation to thermal equilibrium.
Broadly speaking, this term can be considered as the mean-field
correction term in the Vlasov equation due to finite $N$.
The violent relaxation incorporating this term
can be described as follows. In a finite-$N$ system, there
exists the mean-field fluctuations with amplitudes proportional
to $1/\sqrt{N}$ in addition to the mean field from the fluid limit.
This combined mean field 
self-consistently governs the early dynamics altogether
and it deviates the distribution function
$\bar{f}$ from the Vlasov limit
at a rate proportional to $1/N$.
The $N$-dependence arising from the proposed term, which
is of mean-field origin, is unlike the $N$-dependent
collisional effect that involves the
two-point correlation and comes into play much later.
The proposed mechanism is therefore a distinct 
process from the collisional evolution.
Because the violent relaxation time scale
does not depend on $N$, the accumulated deviation at the end
is expected to scale with $1/N$.
In other words, such effect manifests the $N$-dependence 
of the QSS which is not foreseen by the original LB theory.
The $1/N$ term can otherwise be regarded
as the mean-field correction term of the Vlasov equation 
when considering finite-$N$ systems.
As for the collision term, this fluctuation term
vanishes as $N\rightarrow\infty$. In that limit, the evolution
follows strictly the Vlasov equation, i.e.,
$df/dt=d\bar{f}/dt=0$. This happens when one directly solves
the Vlasov equation for the evolution of $f$, either
numerically or semi-analytically
\cite{janin_1971,de_buly+gaspard_2011,colombi+touma_2014,colombi_2015,barnes+ragan_2014,ragan+barnes_2019}.
Note that this estimate of the $1/N$ term
is valid for any long-range interacting systems as long as
the Poissonian fluctuations are present.
Although applying the perturbative analysis to the
Vlasov equation yields the analytical results
that are analogous, in a broad sense, to those
involving the collision term from the BBGKY hierarchy,
the scope of the
applicability of our analysis is more limited.
Most importantly, the violent relaxation, which
is our focus here, leads to the QSS
that differs greatly from the initial state in most of
the cases and the system
evolves rapidly in a short time scale.
The full perturbative analysis therefore fails to
be valid. As a matter of fact, our analysis
only provides an estimate of the rate of the departure
from the Vlasov limit starting from the non-equilibrium
initial condition, via the analysis that yields lastly
Eq. (\ref{dfdt_perturbed}).
We do not provide a full description of how the
perturbation (or $\delta f$) evolves, nor how the
final QSS appears.
On the contrary, the collisional term drives
the QSS slowly and quasi-statically through the series of the
dynamical equilibria in the course of the thermalization
and the system eventually attains thermal equilibrium.

In a confined system such as the HMF model, the term $\delta f$ 
can be assumed to take a form of the standing wave. 
A recent work by \cite{chavanis_2022PhyA} performed the 
quasi-linear analysis in that way and investigated
its effect on the QSS. The results were compared 
with $N$-body simulations and it was found that the analysis
was applicable in a certain range of energy. 
Otherwise, a large deviation from the theory was found.
It may appear that the starting point 
and the purpose of this work are similar to ours:
the perturbative form 
of the distribution function is applied 
to the Vlasov equation and 
its consequence to the QSS is examined, 
but the in-depth details are different.
The major difference is that
they applied the analysis to the regime in which the
QSS did not differ much from the initial state.
In our case, we start from
the far out-of-equilibrium initial state.
Hence, our analysis does not aim to describe the
detailed phase-space density distribution of the QSS
via $\delta f$, but we
only describe how far the global properties of
the phase-space density
of the QSS can deviate from the Vlasov limit.
Another difference is that their work regarded 
the perturbations in the fluid limit
while in our study, those perturbations represent
the local finite-$N$ fluctuations.
Therefore, the resulting QSS from our analysis
is the QSS with the finite-$N$ effect involved
and the test with the $N$-body simulations
is straightforward.

\section{Simulation details, initial condition and units}
\label{ic_unit_sim}

We test the hypothesis of the $1/N$ term that arises from
the above analysis in the $N$-body simulations
of one-dimensional self-gravitating system 
of particles of identical mass $m$. This model
is also known as the self-gravitating sheet model.
The Newtonian gravitational acceleration 
from the particle $j$ located at $x_{j}$
acting on the particle $i$ at $x_{i}$ reads
\begin{equation}
a_{ji} = gm\frac{x_{j}-x_{i}}{|x_{j}-x_{i}|}
\label{force_ij}
\end{equation}
where $g$ is the gravitational constant. 
The force expressed in Eq. (\ref{force_ij}) is 
the mutual attractive constant force regardless of separation.
This facilitates the numerical integrations of the equations of motion 
of particles as they can be carried out in the exact way up to
the crossing where the forces on the encountering
particles have to be redetermined.
The most optimized algorithm to handle this is
the heap-based algorithm in which the crossing times
of all pairs of adjacent particles are stored and managed
in a data heap \cite{noullez_fanelli_aurell_2003_heap}.
The accuracy is such that the deviation  
of the total energy is comparable to the machine error
at each crossing. At the end, the deviation of the total
energy is less than $10^{-6} \%$.

For the initial condition, we adopt 
the rectangular waterbag configuration in phase space 
in which the initial position and velocity of particles 
are randomly distributed in the range of $(-x_{0},x_{0})$ 
and $(-v_{0},v_{0})$, with uniform probability. 
Thus, the uniform initial phase-space density (or $f_{0}$) 
inside the waterbag reads
\begin{equation}
f_{0} = \frac{M}{4x_{0}v_{0}}
\label{virial_b0}
\end{equation}
where $M=mN$ is the total mass and the phase-space density
is zero outside. By this construction, the phase-space
distribution is a two-level function 
which facilitates the analysis as $f_{0}$ 
is the integral of motion and it also indicates
the maximum attainable phase-space density according
to the LB theory. We choose to parameterize  $f_{0}$ by
the initial virial ratio $b_{0}\equiv \frac{2T_{0}}{|U_{0}|}$,
where $T_{0}$ and $U_{0}$ are the initial kinetic and potential
energies of the system, respectively, as 
\begin{equation}
f_{0}=\sqrt{\frac{M}{2gL^{3}b_{0}}}
\label{f0_b0}
\end{equation}
where $L=2x_{0}$ is the initial system size.
Parameterizing the initial state by $b_{0}$ helps to 
identify easily how far the initial state is from
the virial equilibrium. Throughout this study,
we use a system of units where the gravitational constant $g$, 
the total mass $M$,
and the system initial length $L$ are unity. The unit of time
$t_{d}$ is chosen to be 
\begin{equation}
t_{d} = \sqrt{\frac{L}{gM}}
\label{unit_t}
\end{equation}
which corresponds to the free-fall time of the $b_{0}=0$ system.
After all particles are randomly sampled
in phase space following the prescriptions above,
we adjust all positions and velocities by the constants 
so that the center of mass is static at the origin.
Relocating the center of mass in this way is equivalent
to translating the waterbag in phase space while 
the initial degree of randomness, which is our central interest,
is preserved.

Because we study the subtle effect from the random microscopic 
Poissonian fluctuations, which vary between realizations,  
to the macroscopic QSS, the simulation with 
a considerable number of realizations for each 
$b_{0}$ and $N$ is preferable to 
obtain the dependable results. For this reason, 
we simulate $1000, 500, 250, 150, 100$ 
and $50$ different 
realizations for $N=1000, 2000, 4000, 8000, 16000$ and 
$32000$, respectively, and the ensemble average 
can be applied if necessary. 

\section{Numerical results}
\label{nume}

\subsection{Waterbag evolution and phase-space density of QSS}
\label{fe_qss}

\begin{figure*}

\begin{center}
\hspace{-5mm}
\includegraphics[width=18.0cm]{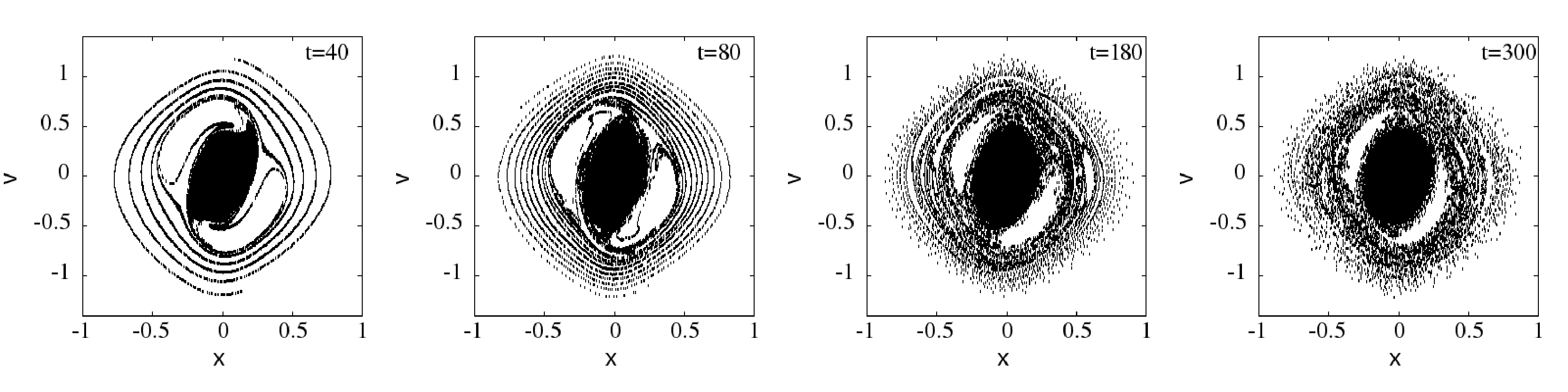}
\end{center}
\caption{Configurations in phase space for $b_{0}=0.1$ and $N=32000$
at different times.}
\label{fig_snap}
\end{figure*}

In this section, we consider first of all the evolution
of the waterbag. Shown in Fig. \ref{fig_snap} is
the evolution of the waterbag in phase space for
$b_{0}=0.1$ and $N=32000$, drawn from a single realization.
The initial waterbag evolves to the QSS which,
as seen at $300 \ t_{d}$, consists of a
dense core and surrounding diluted halo achieved
from wound filaments seen in the snapshots before.
This kind of structure is known as the core-halo structure.
The attainment to this structure can be described
by the Vlasov dynamics as follows.
In brief, the core phase-space density reflects
the density of the initial waterbag, namely $f_{0}$, while the 
filament densities do not. This is because filaments
become finer while they are winding, 
lowering their coarse-grained phase-space density with time
even though the fine-grained density is preserved to $f_{0}$.
On the contrary, the core does not stretch and wind,
so its phase-space density remains at $f_{0}$.
The Vlasov description for the collisionless
relaxation is deemed to be valid until filaments
become discontinuous due to limited resolution from finite $N$
(see \cite{miller_et_al_2023} for review).
We observe that filaments are still intact until $80 \ t_{d}$.
At $180 \ t_{d}$, they start to dissolve, but their trace 
in the halo is still observable.
At $300 \ t_{d}$, filaments are more assimilated.
We also spot a number of persistent holes in the halo, which emerge
between the core and the winding filaments.
The presence of phase-space holes does not significantly
affect the analysis because they are situated well outside
the core, whereas we will examine the core
to test our hypothesis of the $1/N$ term.

\begin{figure}
\begin{center}
\includegraphics[width=8.0cm]{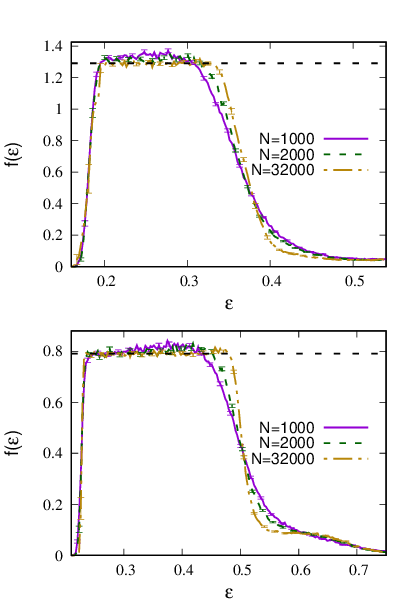}
\end{center}
\caption{Ensemble-averaged phase-space density as a function of 
    the energy per unit mass $f(\epsilon)$ for $b_{0}=0.3$ (top panel)
    and $0.8$ (bottom panel) and 
    for different $N$. Plots are taken at $300 \ t_{d}$.
    Horizontal dashed line indicates the corresponding $f_{0}$.
    Size of error bars corresponds to the
    standard deviation of the mean.}
\label{fig_fe}
\end{figure}

From the visual inspection of the
waterbag evolution, we continue with an inspection
of the phase-space density distribution of the QSS.
Shown in Fig. \ref{fig_fe} is the 
ensemble-averaged phase-space density as a function of the energy 
per unit mass $f(\epsilon)$ at $300 \ t_{d}$ for $b_{0}=0.3$ and $0.8$,
and for various $N$.
To compute $f(\epsilon)$, we first determine the energy distribution
of particles $F(\epsilon)$ such that $\int F(\epsilon) d\epsilon= M$. 
Then, the conversion of the distribution from the energy domain
(or $F(\epsilon)$) to the phase-space domain (or $f(\epsilon)$)
is carried out by the relation
\begin{equation}
F(\epsilon) = g(\epsilon)f(\epsilon)
\label{f_conversion}
\end{equation}
where $g(\epsilon)$ is the density of state 
at energy $\epsilon$ which can be computed from
\begin{equation}
g(\epsilon) = \int_{\Phi_{0}}^{\epsilon}
\frac{2\sqrt{2}}{a(\Phi)\sqrt{\epsilon-\Phi}}d\Phi.
\label{ge_conversion}
\end{equation}
In Eq. (\ref{ge_conversion}), $a(\Phi)=\frac{d\Phi}{dx}$ 
is the magnitude of the acceleration.
The lower integral limit $\Phi_{0}$ corresponds to
the minimum potential energy which is located at $x=0$.
For simplicity, we can set $\Phi_{0}=0$
without loss of generality. We determine the conversion
factor $g(\epsilon)$ by the numerical integration
of Eq. (\ref{ge_conversion}). The ensemble-averaged
$f(\epsilon)$ is obtained by averaging $f(\epsilon)$ from
each individual realization, accompanied with the error
bars estimated from the standard deviation of the mean.
In the plot, we provide $f_{0}$ for comparison.
The advantage of inspecting the energy distribution 
in phase-space domain 
is that it can be directly compared to $f_{0}$ 
which is the conserved quantity in the collisionless evolution
and indicates, in principle, the maximum attainable
phase-space density. Thus, the degeneracy of the system 
and the deviation from that limit can be analyzed.
In addition, the robustness
of the ensemble-averaged $f(\epsilon)$ is
tested by the bootstrapping method and the results
are reported in Appendix \ref{bootstrap_appen}.

Because we aim to inspect the 
subtle properties of the QSS, to ascertain that
the state at the end appropriately represents the 
QSS is crucial. On the one hand, we assure that
the violent relaxation is put to an end well
before that time. According to past studies,
it was demonstrated that the violent relaxation that was
probed by the virial ratio terminated within
$100 \ t_{d}$ \cite{luwel_et_al_1984,mineau_et_al_1990,rouet+feix_1999}.
On the other hand, the fact that the collisional effect,
which also originates from the finite-$N$ fluctuations
in the initial state according to the BBGKY hierarchy, 
is negligible for our conclusion is equally important. 
In a past research work \cite{joyce+worrakitpoonpon_2010},
the collisional relaxation to thermal equilibrium
was tracked by a set of order parameters which were
defined to be zero at the
thermal equilibrium and non-zero at the QSSs.
The evolution of order parameters suggested that
the systems reached the QSS within $100 \ t_{d}$,
in line with other studies, and remained in the
QSSs for a long period of time. 
With $N=800$, the order parameters
saturated well in the QSSs and they did not
start to evolve toward the thermal equilibrium 
earlier than $10^{4} \ t_{d}$.
In this study, we choose to inspect the systems with $N$
higher than that case and the framed time scale is well
before the collisional time scale.
It is worth noting that the BBGKY
hierarchy does not signify the total absence of the 
collisional effect in the time frame considered.
The collisionality is also involved since the start
as with our proposed $1/N$ term.
Nevertheless, we can safely neglect the collisional
effect for our results because, according to past studies,
it becomes significant at much later time.
Another point to be reassured is that,
albeit the presence of persistent phase-space holes 
(see Fig. \ref{fig_snap}), $f(\epsilon)$ is well stationary
in the last $100 \ t_{d}$, which is the time frame of
consideration for further analysis.

From the $f(\epsilon)$ plot, the QSS exhibits
two distinct components: the inner part has $f(\epsilon)$ 
close to $f_{0}$ while the outer part at higher $\epsilon$
has much lower density. This form corresponds to the core-halo
structure as shown in Fig. \ref{fig_snap} and the
violent relaxation leading to such structure can be explained
by the Vlasov equation. An alternative explanation for the
fact that the core ends up in the degenerate state,
which is the state of lowest possible energy, is given
by the theory of the parametric resonance
(see \cite{levin_et_al_2014_pr} for review). 
That theory was initially developed for explaining
the halo formation in an ion system
\cite{gluckstern_1994}, but the adaptation to
self-gravitating systems was also considered
\cite{levin+pakter+teles_2008,levin+pakter+rizzato_2008,teles_levin_pakter_2011,pakter+levin_2011}.
During the violent relaxation,
quasi-periodic oscillation of potential field
caused by a relaxing system can be in resonance
with some particles. As a consequence, those particles
gain the energy and are displaced further away,
constituting the halo.
The energy transfer halts when the core attains 
the degenerate state whose density is limited by $f_{0}$.
The parametric resonance in one-dimensional
system was validated by the oscillation of the
position of a test particle that was in phase with the system
oscillation, so it was able to gain the energy and become part
of halo \cite{teles_levin_pakter_2011,tashiro_2019}.
The fact that the core ends up in the LB degenerate state
has been verified in past works on the self-gravitating sheet model
\cite{lecar+cohen_1971,teles_levin_pakter_2011,joyce+worrakitpoonpon_2011}
and on other long-range interacting systems
\cite{levin+pakter+rizzato_2008,teles+levin+pakter+rizzato_2010},
and our frame of work is based on that fact.
Because the filaments are still intact
at $\sim 80 \ t_{d}$, it is suggested that the parametric resonance
takes place at the later time.

The continuum distribution function of the LB degenerate limit
is the two-level step function given by
\begin{equation}
f(\epsilon)=\begin{cases}
f_{0}, & \text{if $\Phi_{0} < \epsilon < \epsilon_{f}$} \\
0, & \text{otherwise}
\end{cases}
\label{degenerate_lb}
\end{equation}
where $\Phi_{0}$ is the minimum energy and $\epsilon_{F}$ is
the Fermi energy depending on the mass and $f_{0}$.
With a close inspection on the core phase-space density, 
we find that $f(\epsilon)$ can reach the value above $f_{0}$  
and this excess is more noticeable for lower $N$.
If $N$ is higher, the core $f(\epsilon)$
is more uniform and closer to $f_{0}$, albeit with fluctuations,
and $f(\epsilon)$ falls more sharply 
to the halo level beyond the core energy limit.
This implies that the core is closer to the LB
degenerate $f(\epsilon)$.
In fluid limit, the Vlasov equation
suggests that the core distribution function takes
the degenerate form (\ref{degenerate_lb}).
That the deviation from the degenerate distribution
function is more evident if $N$ is lower can be
explained that the fluctuations of mean-field
accelerations are higher in magnitude. This causes larger
deviation of core particle trajectories from Vlasov
mean-field trajectories.
The phase-space density excess in Fig. \ref{fig_fe} 
is out of the scope of the
conventional violent relaxation theory because,
according to the description by that theory, 
$f_{0}$ marks the maximum allowed phase-space density.
We hypothesize that its origin is from the 
additional term in Eq. (\ref{dfdt_perturbed}).
It turns out that such term causes the QSS to be
more concentrated than $f_{0}$ in phase space, specifically
in the core part. Note that our study adopts
a simple waterbag that consists
of particles of identical mass and has a single uniform
density, so a single core emerges.
In a more complicated case, multiple cores
can arise \cite{reidl+miller_1995}.

We remark that the phase-space distribution
for $b_{0}=0.8$ and $N=32000$ exhibits a plateau at
high energy which contradicts the Penrose criterion
from which a single maximum is allowed in the stationary state.
We speculate that
the emergence of this component is attributed to
the parametric resonance that favors a certain range
of the particle energy. These particles 
therefore constitute an overpopulated component at
the high energy range.
Another possibility is that, in continuity
with the plot of phase-space configuration in Fig. \ref{fig_snap},
the underpopulated region is attributed to the voids
that harbor no particles.
Local maxima in the halo
energy range were also spotted in past works
\cite{hohl+feix_1967,yamashiro+gouda+sakagami_1992,yamaguchi_2008}
but this component was found to be meta-stable as
it disappeared when the system reached the thermal equilibrium 
\cite{joyce+worrakitpoonpon_2010}.

\subsection{$N$-dependence of core phase-space density}
\label{deltafe_qss}

\begin{figure}
\begin{center}
\includegraphics[width=8.0cm]{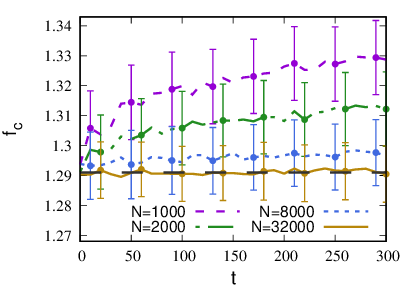}
\vspace{-5mm}
\end{center}
\caption{Time evolution of the ensemble-averaged 
core phase-space density $f_{c}$ for $b_{0}=0.3$
and for various $N$. Size of error bars is the
average of the mean errors in the core region.}
\label{fig_time_fc}
\end{figure}

In continuity with the observed excess of the phase-space
density of the core in Sec. \ref{fe_qss}, we will inspect
it to a greater extent in this section. First of all,
we verify if this excess 
is robust in the QSS and it is really achieved 
from the violent relaxation governed by mean field
with fluctuations. To do so, we define the 
core phase-space density $f_{c}$ to be the averaged
$f(\epsilon)$ in the range of energy where $f > 0.99f_{0}$,
and the time evolution of 
$f_{c}$ for $b_{0}=0.3$ and for various $N$ is 
plotted in Fig. \ref{fig_time_fc}.
We observe that $f_{c}$ increases from $f_{0}$
most prominently in the first $100 \ t_{d}$ 
which is during the violent relaxation.
After $200 \ t_{d}$, $f_{c}$ saturates at the QSS.
A system with low $N$ tends to depart more from 
$f_{0}$ whereas for $N=32000$, $f_{c}$ remains
close to $f_{0}$ all the time.
This plot ascertains, firstly, that the QSS with
the core phase-space density exceeding $f_{0}$  
is a robust structure as a consequence of 
the violent relaxation. 
Secondly, the departure from the theoretical 
limiting value in the fluid limit during the
violent relaxation is more noticeable in
a lower $N$ system. These are in line with the analytical 
result in Sec. \ref{vlasov_analysis}.

\begin{figure}
\begin{center}
\includegraphics[width=8.0cm]{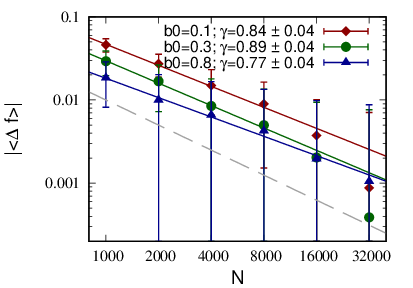}
\vspace{-5mm}
\end{center}
\caption{Magnitude of the averaged $\Delta f $ (or $|<\Delta f>|$)
as a function of $N$ for different $b_{0}$. 
The solid line is the best-fitting $1/N^{\gamma}$ function with
the best-fitting $\gamma$ reported for each case.
The dashed line is the $1/N$ decay for comparison. 
Size of error bars corresponds to
the averaged mean error in that time window.}
\label{fig_deltaf}
\end{figure}

The deviation of $f_{c}$ from $f_{0}$ can be
investigated in a more quantitative way by defining 
\begin{equation}
\Delta f \equiv \frac{f_{c}-f_{0}}{f_{0}}
\label{delta_f_fc}
\end{equation}
to be the phase-space density excess of the core and
the plot of the magnitude of the ensemble-averaged
$\Delta f$ (or $|<\Delta f>|$) for different $b_{0}$
and $N$ is shown in Fig. \ref{fig_deltaf}.
The time average is also performed in
the time window of width $50 \ t_{d}$
from $250-300 \ t_{d}$ to suppress the time fluctuations.
The size of the error bars corresponds to the averaged mean errors
in the same time window.
For each $N$, $|<\Delta f>|$ tends to be higher when $b_{0}$
is lower, which implies that a further out-of-equilibrium
initial state leads to a QSS that
is further away from the Vlasov limit.
A waterbag of lower $b_{0}$ 
collapses to the center more violently
because of lower support from velocity dispersion.
Thus, local fluctuations in the initial state
gather more concentratedly in the core,
leading to a more pronounced
effect from the mean-field fluctuation term.
Despite the variation in magnitude with $b_{0}$,
we observe the decrease of $|<\Delta f>|$
with $N$ for all $b_{0}$. To verify if the decrease is
in accordance with the $1/N$ estimate, we perform the
curve-fitting with the $1/N^{\gamma}$ decay, where
$\gamma$ is a positive real number, for all $b_{0}$ and
the results are depicted in Fig. \ref{fig_deltaf}.
The line designating the $1/N$ decay is also put
for comparison. The curve-fitting with
the decreasing function that is not restricted
to $1/N$ allows us to test the validity of the
$1/N$ estimate properly and, if it is about
to deviate from that estimate, to evaluate
how large the discrepancy is.
In all cases, the best-fitting $\gamma$ is not
far from $1$. This affirms 
the reasonable applicability of the $1/N$ fluctuation term in
the perturbed Vlasov equation
that we proposed in Sec. \ref{vlasov_analysis}.
While it is desirable if we keep increasing $N$ and inspect
further this finite-$N$ effect, we foresee some limitations
by doing so. For $N=32000$, the deviation is of the order of a
part per thousand or even below, and the deviation
should be weaker if $N$ is higher.
Therefore, the capability to measure it properly is doubtful.

In past studies, the disagreement between
the theoretical prediction of the QSS
based on the continuum limit and 
the $N$-body simulations has  
been remarked in the HMF model (see, e.g.,
\cite{yamaguchi+barre+bouchet+dauxois+ruffo_2004,antoniazzi+califano+fanelli+ruffo_2007}).
Because that model is confined in a finite spatial domain, 
the stability analysis of the continuum 
distribution functions for
the most probable state as the QSS is feasible. 
In our case, because a complete stability 
analysis to obtain the prediction of the QSS
is not simple, we choose
to infer from past studies that the core of the QSS
follows the LB degenerate state. By our choice of 
initial conditions, the degenerate distribution function
has a constant phase-space density equal to $f_{0}$ and 
we adopt it as the expected value. We not only find
the discrepancy between the theory and the simulations
as the past studies of the HMF model did, but we also 
attempt to understand better the origin of that
discrepancy and its $N$-dependence. We are able to 
explain the cause of that deviation by the hypothesis of
the $1/N$ term. If this $1/N$ term is generic 
for any long-range interacting systems,
we speculate that the deviation of 
the measured parameters in the $N$-body system
from the prediction in the fluid limit might scale
with $N$ in the same way as our result.

\subsection{Core density profile and its agreement with LB theory}
\label{density_qss}

\begin{figure}
\begin{center}
\includegraphics[width=8.5cm]{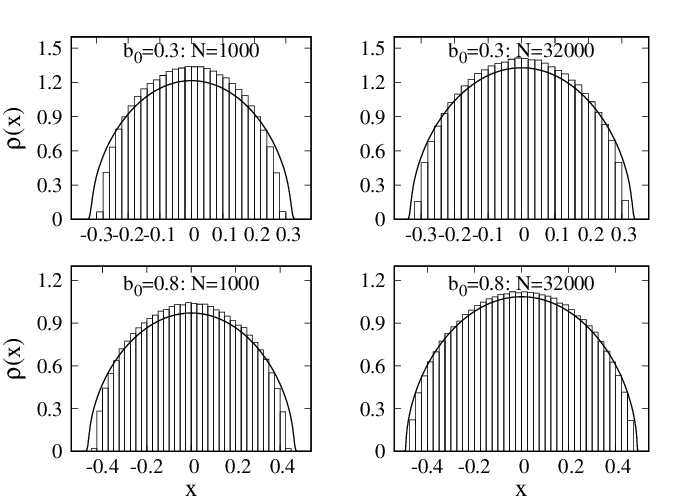}
\vspace{-5mm}
\end{center}
\caption{Density profile of the core $\rho(x)$
for $b_{0}=0.3$ (two top panels) and $0.8$ (two bottom panels)
in histogram.
For each $b_{0}$, we depict $\rho(x)$ for $N=1000$ and $32000$.
Solid line is the averaged density profile of the
LB degenerate limit.}
\label{fig_rho}
\end{figure}

To inspect the core in more details, the ensemble-averaged
density profile of the core $\rho (x)$ is plotted in
Fig. \ref{fig_rho} for some selected $b_{0}$ and $N$.
We define the core to be the component below $\epsilon_{c}$
at which $f(\epsilon)$ falls to $0.99f_{0}$.
We provide the corresponding density profile of the
LB degenerate limit for comparison, which is numerically
calculated from the distribution function (\ref{degenerate_lb})
via the Poisson equation. The plot of the LB density profile
in Fig. \ref{fig_rho} is obtained by averaging 
the LB profiles calculated from  
the core mass of each realization.
The bootstrapping test for the robustness of the
core density profile is also performed and
the results are reported in Appendix \ref{bootstrap_appen}.
From the plot, the deviation from the LB degenerate limit
is remarkable when one inspects the density profile
although $|<\Delta f>|$ is as small as a few percents
or sub-percent. For $b_{0}=0.3$, the simulated
core is evidently more concentrated than the
degenerate limit counterpart.
These two profiles become closer to each other
if $N$ increases from $1000$ to $32000$. 
For $b_{0}=0.8$, the over-dense core is also observed for
$N=1000$ but the discrepancy is not as large as in 
the $b_{0}=0.3$ plot for the same $N$. 
As $N$ reaches $32000$, 
the simulated and theoretical profiles almost coincide.
That the core density profile becomes closer to the LB
degenerate limit for higher $N$ is in coherence
with the $f(\epsilon)$ plots in Fig. \ref{fig_fe}.
The plots in Fig. \ref{fig_rho} are another evidences
of the influence of
the proposed $1/N$ term that deviates the core from
the continuum degenerate limit.
In past studies, a reasonable agreement between
the simulated QSS and the LB profile has been captured
for $b_{0}$ close to $1$
\cite{yamaguchi_2008,joyce+worrakitpoonpon_2011}.
In this work, we prove further that the agreement 
with the LB theory, albeit delimited to the core, 
improves if we increase $N$. That tendency 
is found even if $b_{0}$ is considerably far from $1$.
Note that a better agreement for a higher $N$ system
may not be realized if we consider the entire system
because the situation is not simple.
This is because the underlying parametric resonance
that brings the core to the degenerate state,
on the other hand, decouples the halo from the core.
Thus, the entire system is not fully mixed.
This scenario is recognized as the incomplete relaxation
which prevents the QSS to take the LB form \cite{chavanis_2006PhyA}.
Another possibility of the incomplete
relaxation is that the systems consist of particles
of different masses. This leads to the mass segregation
which also prevents the system to
be fully relaxed \cite{yawn+miller_1997,yawn+miller_2003}.

\begin{figure}
\begin{center}
\includegraphics[width=8.5cm]{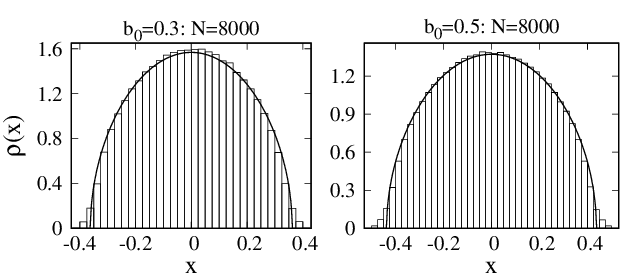}
\end{center}
\caption{Density profile of the central component $\rho(x)$ 
below the energy $\epsilon_{c,10}$ 
at which $f(\epsilon)$ falls to $0.1f_{0}$ for
$b_{0}=0.3$ (left panel) and $0.5$ (right panel),
both of which have $8000$ particles.
Solid line represents the density profile
of the LB degenerate limit for the corresponding $f_{0}$
and the mass of this component.}
\label{fig_rho10}
\end{figure}

Finally, we revisit the ansatz of Teles et al (2011) 
\cite{teles_levin_pakter_2011} which proposed 
that the core-halo structure can be 
well fitted by the three-level distribution function.
In that ansatz, the core phase-space density is fixed to 
$f_{0}$ while the constant halo phase-space density,
the core energy limit and the halo energy limit 
are determined self-consistently by the mass and energy
constraints, and the energy of the furthest particle.
Otherwise, the distribution function is zero.
Our results make evident that the simulated $f(\epsilon)$ 
is more complex than the three-level configuration 
(see Fig. \ref{fig_fe}). 
We note the intermediate region where $f(\epsilon)$
decreases from the core density to the halo density
that spans a considerable range of $\epsilon$,
and we also note that the profile of this region
varies considerably with $N$.
Furthermore, we have inspected the density profile of
the core that is restricted to be within 
the near-degenerate
central region and a poor agreement with 
the LB degenerate
distribution function has been observed in most cases.
It is nevertheless worth testing that ansatz
for the simulated core whether the agreement improves
if we escalate the core energy limit
from the value adopted in past sections while 
the core phase-space density is still fixed to $f_{0}$.
To do this, we shift the core cutoff energy
to the intermediate region at which $f(\epsilon)$ 
falls to $0.1f_{0}$, i.e., $\epsilon_{c,10}$,
in the attempt to average the over-dense core and 
a part of less dense intermediate region.
The density profile of the core which is now 
constituted from the particles with energies below 
$\epsilon_{c,10}$
and the corresponding LB degenerate profile 
are plotted in Fig. \ref{fig_rho10} for $b_{0}=0.3$ and 
$0.5$, both of which comprise $8000$ particles 
that are comparable to the original work 
of Teles et al (2011) \cite{teles_levin_pakter_2011}.
We observe a good agreement with the LB degenerate
density profiles of these central components.
Although we do not investigate further the 
full core-halo distribution function,
our result affirms the applicability of
the two-level function, with modified energy limit,
to the central component.
The agreement found in the $b_{0}=0.5$ plot 
is in accordance with \cite{teles_levin_pakter_2011}
whereas we prove that the agreement 
can be found even though $b_{0}$ is significantly below.
We note a density tail beyond the
LB profile and it is the part that exhibits
an obvious inconsistency. We speculate that it originates
from the fluctuations of the energies of particles
near the cutoff energy as the particles merely inside
the degenerate profile can possibly have
the energies higher than the cutoff energy,
which are then excluded from the consideration,
and vice versa for the particles merely outside.
Despite this good agreement with numerical results, 
we may have a different opinion because
the importance of the 
finite-$N$ effect might be overlooked.
We demonstrate that such effect causes 
the evident variation with $N$ of the core.
This implies the violation of the collisionless 
Vlasov equation due to the 
finite-$N$ effect that should be of concern.

\section{Conclusion}
\label{conclusion}
We have investigated the significance of
the local finite-$N$ fluctuations embedded in the
initial states of systems of particles
to the QSS that marks the end of a violent relaxation.
Based on the conventional mean-field description of 
the violent relaxation, the particle number $N$ has 
no influence on either the violent relaxation 
time scale or the statistical description of the QSS.
We have identified the finite-$N$ contribution in
the Vlasov equation as the local mean-field
fluctuation term proportional to $1/N$. 
This term is  
intrinsically different from the collision term
since it is of mean-field origin unlike the collision
term that involves the two-body interaction.
Therefore, this term is effective in the violent
relaxation in the way that it deviates the system
from the Vlasov limit at a rate proportional to $1/N$.
This implies the $N$-dependent QSS in the way that
the deviation degree from the Vlasov limit is scaled by $1/N$.

That hypothesis is tested in the simulations of one-dimensional
self-gravitating systems starting from a simple waterbag initial 
condition of constant phase-space density $f_{0}$.
The phase-space density distribution
is a two-level function: it is equal to $f_{0}$ inside
the waterbag and $0$ elsewhere. In this way,
$f_{0}$ can serve as a reference of the degeneracy.
We focus on the core of the QSS which, 
according to past studies, typically attains 
the LB degenerate state as a result of the 
parametric resonance during 
the violent relaxation. 
The LB degenerate limit designates the state 
of lowest possible energy in the fluid limit and
it is unique to $f_{0}$ and mass.
We find that the core phase-space density
can exceed $f_{0}$ which marks the maximum attainable
value according to the conventional violent relaxation theory. 
By a proper measurement of the phase-space density
excess, we find that the deviation is 
in agreement with the $1/N$ estimate.  
A larger deviation from the Vlasov limit in a system of lower $N$
is further affirmed by the inspection of the 
core density profile where the agreement between the
simulated core density profile and the
LB degenerate density profile improves for higher $N$. 
In conclusion, we underline the importance of the correction term
in the Vlasov equation when dealing with $N$-body systems.
Although our study provides a reasonably precise description
of the $1/N$ term that is applicable to the core, the detail
of the entire QSS remains to be resolved.

\begin{acknowledgments}
This research has received funding support from the 
National Science, Research and Innovation Fund (NSRF) 
via the Program Management Unit for Human Resources 
\& Institutional Development, Research and Innovation
(PMUB) (grant number B05F640075), and partly by
Thailand Science Research and Innovation (TSRI)
via Suranaree University of Technology (SUT) 
(grant number 179349). Numerical simulations
are facilitated by HPC resources of the National
Astronomical Research Institute of Thailand (NARIT).
The author also thanks useful comments from anonymous reviewers
and C. Herold in improving the manuscript.
\end{acknowledgments}

\bibliographystyle{apsrev4-2.bst}

%

\appendix

\section{Bootstrapping test for the QSS}
\label{bootstrap_appen}

\begin{figure*}
\begin{center}
\includegraphics[width=14.0cm]{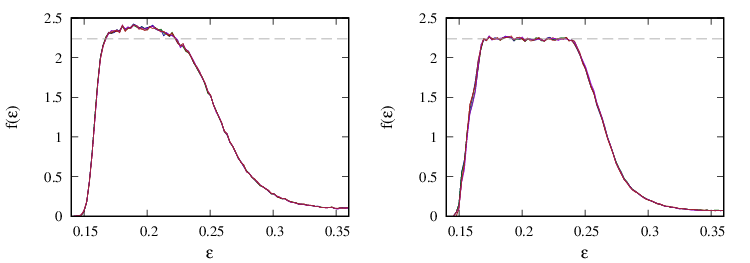}
\end{center}
\caption{Phase-space density distribution $f(\epsilon)$ for
$b_{0}=0.1$ and $N=1000$ (left panel) and
$b_{0}=0.1$ and $N=32000$ (right panel) for 5 different
subsamples, each of which consists of randomly chosen
$80\%$ of realizations. The horizontal dashed line
indicates the value of $f_{0}$.}
\label{fig_fe_boot}
\end{figure*}

\begin{figure*}
\begin{center}
\includegraphics[width=14.0cm]{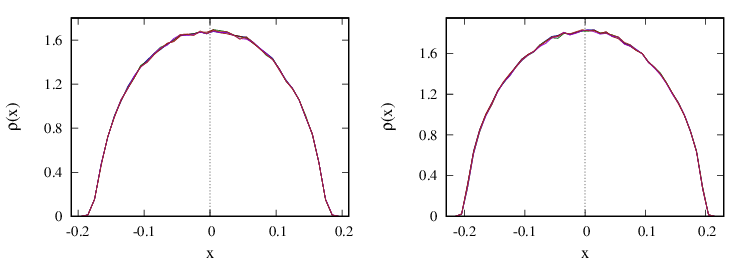}
\end{center}
\caption{Density profile of the core $\rho(x)$ for
$b_{0}=0.1$ and $N=1000$ (left panel) and
$b_{0}=0.1$ and $N=32000$ (right panel) for 5 different
subsamples, each of which consists of randomly chosen
$80\%$ of realizations.}
\label{fig_rho10_boot}
\end{figure*}

In this appendix, we report the results of the bootstrapping
test for the robustness of the QSS. We choose the cases with $b_{0}=0.1$
and $N=1000$, which is furthest away from
the Vlasov limit, and $b_{0}=0.1$ and $N=32000$, which 
has the lowest number of realizations. We compute 
the phase-space density distribution $f(\epsilon)$
(Fig. \ref{fig_fe_boot}) and the
density profile of the core $\rho (x)$ (Fig. \ref{fig_rho10_boot})
for 5 different subsamples.
The horizontal line denoting $f_{0}$ is also provided in
the $f(\epsilon)$ plots. 
In each subsample, $20\%$ of the realizations are randomly
pulled off. Although the variation of the QSS among realizations
is expected, especially for cases with low $N$, both figures
ascertain that the number of realizations for our study
is sufficient to yield the dependable results as
the computed profiles almost coincide.

\end{document}